\def\wordcount {1}		% Use for word count

\ifdefined\preprint
  \documentclass[preprint,review,12pt]{elsarticle}
\fi
\ifdefined\wordcount
  \documentclass[final,3p,times,twocolumn]{elsarticle}
\fi
\ifdefined\final
  \documentclass[final,3p,times,twocolumn]{elsarticle}
\fi

\usepackage{graphicx,stfloats}
\usepackage{color}

\usepackage{subfigure}
\usepackage{amssymb}
\usepackage{amsthm}
\usepackage{textgreek}
\usepackage{amsmath}

\biboptions{sort&compress}

\journal{Proceedings of the Combustion Institute}

\begin{document}

\newcommand{\chem}[1]{\ensuremath{\mathrm{#1}}}
\newcommand{\pa}[2]{\frac{\partial #1}{\partial #2}}
\newcommand{\cBar}{\ensuremath{\left\langle c \right\rangle}}

\begin{frontmatter}

\title{Flame- and flow-conditioned vorticity transport in premixed swirl combustion}

\author{Askar Kazbekov}
\author{Adam M. Steinberg\corref{cor1}}
\ead{adam.steinberg@gatech.edu}

\address{Daniel Guggenheim School of Aerospace Engineering, Georgia Institute of Technology, Atlanta 30332, USA}
\cortext[cor1]{Corresponding author:}

\begin{abstract}
This paper presents an experimental analysis of flame-induced enstrophy transport in premixed swirl combustion at Karlovitz numbers between 20-50. Such flames posses a large-scale pressure field that -- in addition to the pressure fields associated with small-scale turbulent vortices -- can interact with density gradients to produce baroclinic torque. Simultaneous tomographic particle image velocimetry and formaldehyde planar laser induced fluorescence measurements are used to obtain high-resolution velocity and progress-variable fields. This allows statistical evaluation of the various terms in the enstrophy transport equation. The impact of small- and large-scale pressure gradients is assessed by conditioning the baroclinic torque on the position of the fluid within the instantaneous flame front, within the flame brush, and axially within the combustor. At all conditions studied, the baroclinic torque was a significant contributor to enstrophy transport, with a comparable magnitude to vortex stretching and viscous diffusion. Enstrophy attenuation and production by baroclinic torque tended to occur towards the reactant and product sides of the instantaneous flame surface, respectively. However, the value of the baroclinic torque also depended equally strongly on the position in the combustor. Hence, both small- and large-scale pressure fields can result in significant enstrophy changes through baroclinic torque. This is evidence both that flame-induced vorticity dynamics are significant in swirl combustion, and that large-scale geometry-dependent flow fields can impact flame-generated turbulence.

%The sources of mean baroclinic torque enstrophy production and attenuation are examined in a turbulent premixed swirl-stabilized flame. Baroclinic torque and progress variable fields were experimentally measured using  at $\mathrm{Re_j}=26,000-46,000$. The baroclinic torque is conditioned on instantaneous and mean progress variables, and axial position to infer the source of baroclinicity from the multiple sources of pressure gradients in the swirling flow. The magnitude of baroclinic torque enstrophy attenuation and production was found to strongly depend on the position of the flame inside the flame brush. Predominant enstrophy production was found at high $c$ and $\cBar$, whereas attenuation was dominant at earlier stages of combustion.
 
\end{abstract}

\begin{keyword}

Flame-Generated Vorticity \sep Turbulent Combustion \sep Laser Diagnostics \sep Swirl Flames

\end{keyword}

\end{frontmatter}

%% Please do not modify the following three lines
\ifdefined \wordcount
\clearpage
\fi

\section{Introduction}
\label{Introduction}
Premixed flames both influence and are influenced-by turbulence. The turbulence characteristics at scales around and below those of the laminar flame thermal thickness ($\delta_\mathrm{L}^0$) alter scalar gradients -- and hence diffusion of species and enthalpy -- which affect the turbulent burning rate~\cite{Driscoll2008, Lipatnikov2005, Lipatnikov2018}. Furthermore, the flame influences the turbulence in the products, which therefore may exhibit different fluctuation intensities and greater anisotropy (due to the anisotropic nature of premixed flames) than in the absence of combustion~\cite{Furukawa2002,Bobbitt2016a, Hamlington2011, Chakraborty2016}. 

Because vorticity ($\vec{\omega}$) concentrates at the smallest scales of a turbulent flow (i.e the Kolmogorov scale $\eta$) and many practical turbulent flames have Karlovitz numbers of $Ka \propto \left(\delta_\mathrm{L}^0/\eta\right)^2 = \mathcal{O}(10)$, vorticity dynamics commonly are used to articulate the flame-scale effects of combustion on turbulence~\cite{Hamlington2011, Lipatnikov2019, Chakraborty2016, Renard2000}. Changes in the vorticity magnitude can be described by the enstrophy ($\Omega = 1/2\omega_i\omega_i$) field, which follows
\begin{align}
    \frac{1}{2}\underbrace{\left(\pa{\Omega}{t}+u_i\pa{\Omega}{x_i}\right)}_L = &\underbrace{\omega_{i}\omega_{j}\pa{u_i}{x_j}}_{I} -
    \underbrace{\Omega \pa{u_k}{x_k}}_{II}+  \underbrace{\frac{1}{\rho^2}\omega_{i}\varepsilon_{ijk}\pa{\rho}{x_{j}}\pa{p}{x_{k}}}_{III} \nonumber \\  &+\underbrace{\omega_{i}\varepsilon_{ijk}\pa{}{x_j}\left(\frac{1}{\rho}\pa{\tau_{km}}{x_m}\right)}_{IV}
    \label{e:enstrophyTransport}
\end{align}
where $\varepsilon_{ijk}$ is the cyclic permutation tensor, and $\tau_{ij}$ is the viscous stress tensor. In Eq.~\ref{e:enstrophyTransport}, the material derivative ($L$) is affected by vortex-stretching ($I$), dilatation ($II$), baroclinic torque ($III$), and viscous diffusion ($IV$). 

In constant-density flows, enstrophy dynamics are controlled by the competition between vortex stretching and viscous diffusion. While flames affect both of these processes~\cite{Kolla2014, Kolla2016}, the fundamentally new processes that become active in flames are the dilatation and baroclinic torque. Since $\partial_k u_k>0$ in low Mach number flames, dilation acts purely as an enstrophy sink. In contrast, the baroclinic torque may act as a source or sink depending on the relative alignment of the pressure- and density-gradients. Hence, flame-induced vorticity generation is attributed to baroclinic torque~\cite{Louch1998, Mueller1998, Sinibaldi1998}.

Early studies of laminar flame-vortex interactions demonstrated attenuation and generation of vorticity through the flame, with the product-side vorticity having opposite rotation to the original reactant-side vorticity \cite{Mueller1998, Louch1998, Sinibaldi1998}. Similar observations were made during flame/vortex interactions in weakly turbulent Bunsen flames, which indicated that flame-generated vorticity could mitigate flame wrinkling \cite{Steinberg2008}. Direct numerical simulations (DNS) of freely propagating statistically 1D flames in weak turbulence also have indicated that vorticity generated by baroclinic torque can decrease the flame area \cite{Lipatnikov2014, Lipatnikov2019}. This is in contrast to studies that attribute self-acceleration of flames to flame-generated vorticity, and demonstrates the myriad of effects that may be induced by flame/turbulence coupling \cite{Karlovitz1951, Scurlock1953}.

The aforementioned DNS configuration of 1D freely propagating flames has been used to describe the impact of combustion on vorticity across a range of conditions~\cite{Louch1998, Mueller1998}. At low Mach numbers, the pressure gradients responsible for baroclinic torque arise primarily from the vortices themselves and scale as $\nabla p \propto \rho u_\mathrm{\eta}^2/\eta$, where $u_\eta$ is the Kolmogorov velocity scale~\cite{Bobbitt2016}. In this situation, baroclinic torque (and dilation) becomes decreasingly significant relative to vortex stretching and viscous diffusion with increasing $Ka$ \cite{Bobbitt2016,Dopazo2017, Chakraborty2016}. The vorticity dynamics therefore approach those of non-reacting turbulence, with the caveat of increased dissipation in the products due to the higher viscosity. 

However, many practical combustion applications involve flames in flows with mean pressure gradients. For example, gas turbine engines generally employ swirl flames, wherein the adverse axial pressure gradient is responsible for formation of the central recirculation zone (CRZ) necessary for flame stabilization. The pressure gradients associated with baroclinic torque therefore arise from a combination of large-scale flow/geometric parameters (generally scaling as $\rho U^2/L$, where $U$ is a measure of the bulk fluid speed and $L$ is a characteristic combustor dimension) and small-scale turbulent fluctuations. 

We recently demonstrated that the relative significance of baroclinic torque (compared to vortex stretching and viscous diffusion) on enstrophy transport remained roughly constant with increasing $U$ in a set of turbulent premixed swirl flames. Other studies of flames in variable area ducts~\cite{Geikie2018} and high Mach numbers~\cite{Chambers2017a, Poludnenko2011} also indicated the importance of system-scale pressure gradients on enstrophy transport. The objective of this paper is to further articulate the impacts of small- and large-scale pressure gradients on flame-induced vorticity transport in swirl flames by conditioning the terms in Eq.~\ref{e:enstrophyTransport} on aspects of the local flame and global flow field.

\section{Experimental Setup}

Experiments were performed using simultaneous tomographic particle image velocimetry (TPIV) and \chem{CH_2O} planar laser induced fluorescence (PLIF) in three unconfined premixed swirl flames. The experimental setup was identical to that presented in our previous work~\cite{Kazbekov2019}; only a brief summary is given here.

\begin{figure}[!t]
    \centering
    \subfigure[Swirl burner with with measurement locations. The inset shows the mean flow streamlines overlapping the mean progress-variable field at $z=0$~mm for Case~3.]{\includegraphics[width=67 mm]{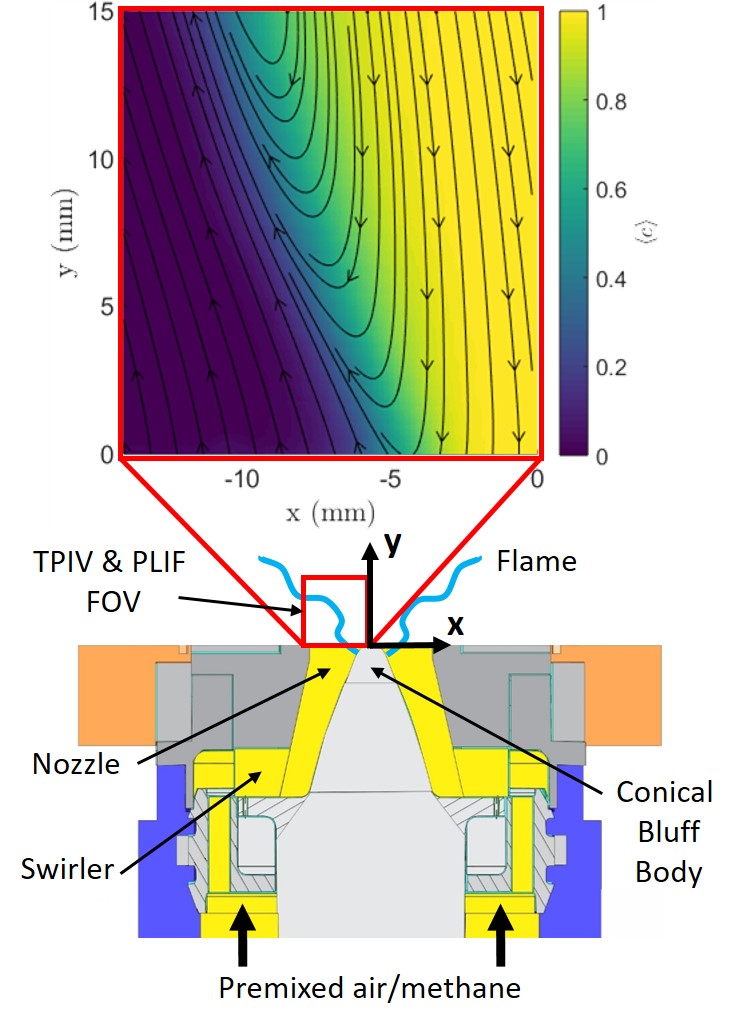}}\\
    \subfigure[Diagnostics configuration]{\includegraphics[trim = {30 0 30 0},clip,width= 67 mm]{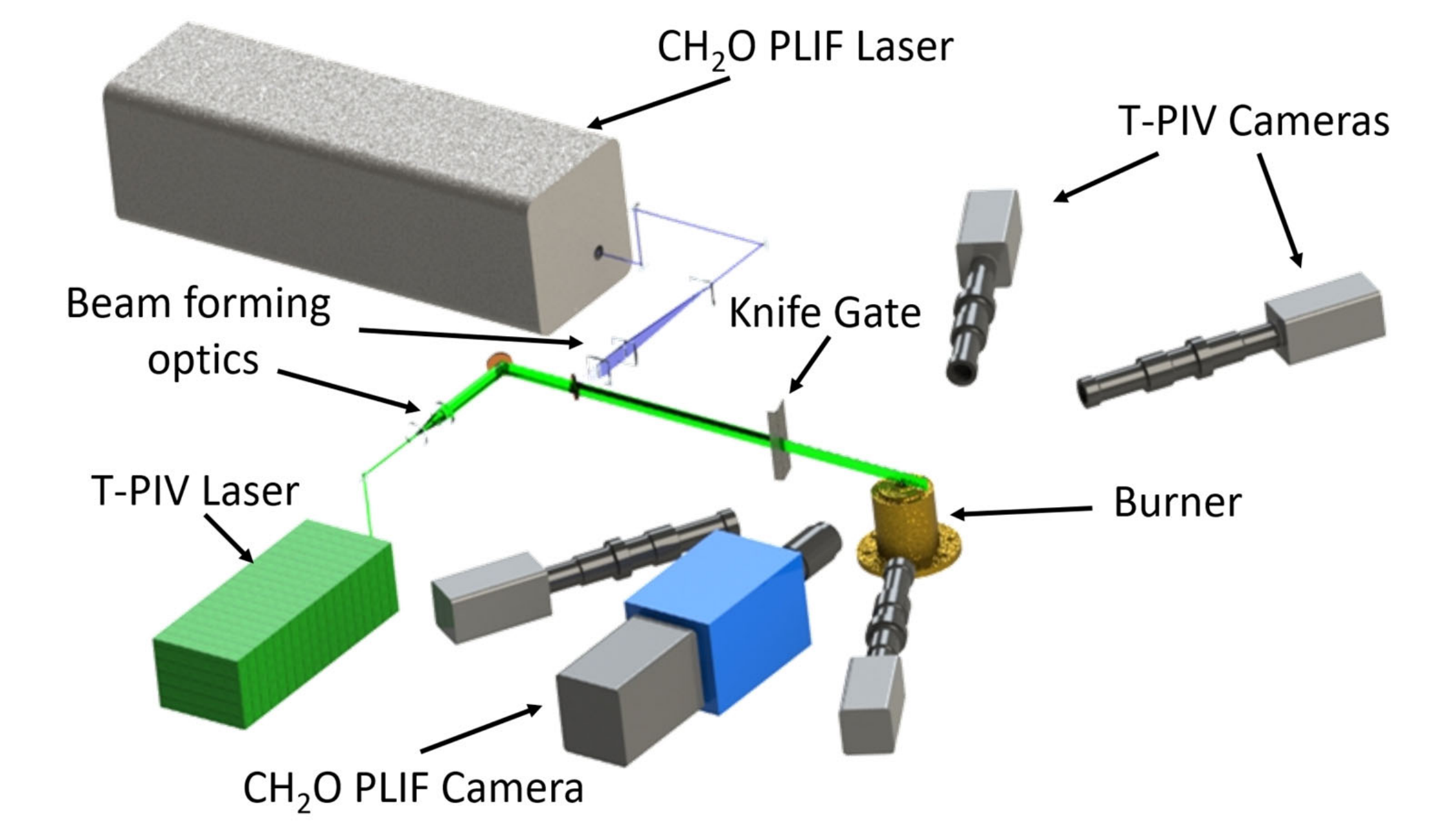}}
    \caption{Experimental configuration.}
    \label{f:setup}
\end{figure}

Flames were stabilized using a gas turbine model combustor (Fig.~\ref{f:setup}(a)) that is identical to that originally described by Meier et al.~\cite{Meier2007}, but with the combustion chamber removed to prevent seed deposition on the windows from interfering with the measurements. Air and fuel (methane) flow rates were metered using mass flow controllers (Brooks, 1\% full-scale uncertainty); premixed well upstream of the combustor plenum to achieve an equivalence ratio of $\phi=0.85$; passed through a radial swirler (measured swirl number of 0.55 at nozzle exit); and expelled through a 27.85~mm diameter nozzle having a conical bluff body along the centerline. 

The bulk flow rate ($U$, volumetric flow rate divided by nozzle exit area), was varied to achieve the conditions shown in Table~\ref{t:testConditions}. Here, the Karlovitz number was calculated as $\mathrm{Ka} = (u^\prime/s_\mathrm{L}^0)^{3/2}(\delta_\mathrm{L}^0/\ell)^{1/2}$, where $u^\prime$ is the root-mean-squared velocity fluctuations in the shear layer between the inflowing reactants and CRZ for the non-reacting flow and $\ell$ is the integral length scale, which was taken to be the thickness of the shear layer. The laminar flame speed ($s_\mathrm{L}^0 = 30.7$~cm/s) and unstretched flame thickness ($\delta_\mathrm{L}^0=0.49$~mm) were calculated using the freely propagating flame model in Cantera~\cite{Cantera}. The turbuelnce Reynolds number was calculated as $\mathrm{Re_T}=u^\prime\ell/\nu_\mathrm{r}$, where $\nu_\mathrm{r}$ is the kinematic viscosity in the reactants.

%\begin{figure}[!t]
%    \centering
%    \includegraphics[width = 67 mm]{Experimental_setup}
%    \caption{Experimental configuration showing the swirl burner with test location. The inset shows the mean flow streamlines overlapping the mean progress-variable field at $z=0$~mm for Case~3.}
%    \vspace{-5 mm}
%    \label{f:setup}
%\end{figure} 

\begin{table}[!t]
\caption{\label{t:testConditions} Test conditions and turbulence properties for each case}
\centering
\begin{tabular}{cccccccc}
Case & $U$ (m/s) & $\mathrm{Ka}$ & $\mathrm{Re_j}$ & $\mathrm{Re_T}$\\\hline
1 & 13.9 & 20 & 26,000 & 1,800\\
2 & 19.3 & 35 & 36,000 & 2,600\\
3 & 24.9 & 50 & 46,000 & 3,300\\
\end{tabular}
\end{table}

%Bulk flow velocity ($U$, ratio of volumetric flow rate to nozzle exit area, with a diameter of 27.85~mm) is varied to achieve turbulence intensities in the range of $\mathrm{Re}_T=2,200 - 2,600$. The jet Reynolds number $\mathrm{Re_j}=UD_e/\nu_r$ spans the range $26,000 - 46,000$. The test conditions are summarized in Table~\ref{t:testConditions}.

All measurements were performed directly above the nozzle, capturing the left branch of the axisymmetric flame (Fig.~\ref{f:setup}(a)). A total of 1,000 simultaneous TPIV and PLIF measurements were made at each test condition. The diagnosics configuration is shown in Fig..~\ref{f:setup}(b).

The TPIV system consisted of a Nd:YAG laser (Quantel Evergreen 200, 532~nm, 10~Hz, 200~mJ/pulse) and four sCMOS cameras (Andor Zyla 5.5, $2048\times2048$ pixels, 6.5~\textmu{m} pixel length). The laser beam was formed into a colimated rectangular slab (width of 1.3~mm) using a series on lenses and a knife edge gate. The cameras were positioned on either side of the laser sheet, at angles between $20^\circ$ and $30^\circ$ relative to laser propagation direction. Each camera was equipped with a 532~nm bandpass filter, long-distance microscope (Infinity K2, CF-1/B objectives, $f/ \# =38$), and Scheimpflug mount (LaVision) to enable off-axis imaging. The flow was seeded with aluminum oxide particles having a nominal diameter of 0.3~\textmu{m}, which resulted in a Stokes number (relative to the Kolmogorov time scale) less than 0.03 in all cases. Particle tomograms and velocity vectors were computed in a commercial software (LaVision DaVis 8.4). Careful optimization of the seed density, laser, imaging, and processing yielded a final resolution of 250~\textmu{m} with 50\% vector overlap (125~\textmu{m} vector spacing) over a $16\times16\times1.3$~mm region.

The \chem{CH_2O} PLIF system consisted of an Nd:YAG laser (Quanta-Ray INDI-40-10, 355~nm, 10 Hz, 70~mJ/pulse), sCMOS camera (Andor NEO 5.5,, $2048\times2048$ pixels, 6.5~\textmu{m} pixel length), image intensifier (LaVision IRO, gate = 100~ns), and camera lens (Tamron, $f=180$~mm, $f/ \# =2.8$), and specialized filter (Semrock FF01-CH2O-50.8-D). Measurements were taken with a field of view of $20\times20$~mm, overlapping the $x-y$ domain at $z=0$~mm (the center of the TPIV volume). The \chem{CH_2O} signal represents the region from the upstream edge of the flame to the location of rapid exothermic reactions.

\section{Analysis, data processing, and uncertainty}
\subsection{Computation of baroclinic torque}
While it is not possible to directly measure the pressure and density gradients involved in the baroclinic torque (term $III$ in Eq.~\ref{e:enstrophyTransport}), the ensemble mean ($\left\langle\cdot\right\rangle$) baroclinic torque can be computed as:
\begin{equation}
    \left\langle III\right\rangle = \frac{1}{2}\left\langle L\right\rangle - \left\langle I\right\rangle + \left\langle II\right\rangle - \left\langle IV\right\rangle 
    %\frac{1}{2}\overline{L} = \overline{I} - \overline{II} + \overline{III} + \overline{IV}
    \label{e:RAETE}
\end{equation} 
The terms $\left\langle L\right\rangle$, $\left\langle I\right\rangle$, and $\left\langle II\right\rangle$ are readily computed from the T-PIV data, recognizing that $\left\langle \partial_t\Omega\right\rangle=0$ in this steady flow. The viscous term $\left\langle IV\right\rangle$ can be estimated using the TPIV and PLIF by invoking a flamelet assumption, as described in Section~\ref{s:progress_var}.

The objective of this analysis is to demarcate the effects of the geometry-dependent large-scale pressure field from those associated with the local turbulence-flame interactions. In planar flames, local enstrophy transport depends on the position through the instantaneous flame surface, which can be characterized by an instantaneous progress variable, $c$ (see Section~\ref{s:progress_var}). The large-scale effects may be characterized by the position of the fluid through the mean flame brush, as given by the mean progress variable field $\left\langle c \right \rangle$ and/or axial position ($y$). We therefore condition the reported results on $c$ and $\left\langle c \right \rangle$, and $y$.

%, and also may depend on the axial position in the combustor ($y$).

\subsection{Progress-variable field and $IV$ measurement}
\label{s:progress_var}
Instantaneous progress-variable fields are necessary to condition the enstrophy transport budgets and, as described below, to compute $IV$. Hence, $c$-fields were estimated using the \chem{CH_2O} PLIF measurements. The distribution of combustion radicals, e.g. \chem{CH_2O} and \chem{OH}, commonly is used to estimate $c$ based on an infinitely thin flame assumption, e.g. \cite{Barlow2007}. Here, \chem{CH_2O} PLIF is used because formaldehyde typically occupies what is classically labeled as the ``preheat zone'' of a laminar premixed flame. We note that Raman scattering measurements by \citet{Dem2014} in flames similar to those studied here indicate that the thermo-chemical structure remains close to that of a laminar flame.

%Formaldehyde radicals are created in the initial stages of hydrocarbon fuel breakdown and are rapidly consumed by exothermic reactions \cite{Temme2015}. For high activation energy reactions, such as \chem{CH_4}/air, the laminar flame thickness is adequately represented by the size of ``preheat zone''. Additionally, the preheat zone is more susceptible to broadening than the thin reaction zone in the presence of finer turbulence structures \cite{Wabel2017, Wabel2017b}. Thus, using \chem{CH_2O} as a proxy for the reaction progress is deemed appropriate.

%\begin{figure}[hb]
%    \centering
%    \includegraphics[width=67mm]{vel_cbar}
%    \caption{A $\cBar$ field overlaid with mean velocity along the central plane of TPIV volume ($z=0$~mm) for Case~3.}
%    \label{f:vel_cbar}
%\end{figure}

Several methods of estimating $c$ from the \chem{CH_2O} PLIF signal are discussed by \citet{Kazbekov2019}, including assuming an infinitely thin flame \cite{Bray1985,Libby1985} with $c=1$ downstream of the \chem{CH_2O}; a three-value approach with $c=0$ upstream of the \chem{CH_2O}, $c=0.5$ in the \chem{CH_2O}, and $c=1$ downstream of the \chem{CH_2O}; and based on linearly interpolating between the two bounding edges of the \chem{CH_2O}-containing region, with the reactant edge corresponding to $c=0$ and the product edge corresponding to $c=1$. The results were found to be qualitatively independent of the method of calculating $c$, with the latter two methods producing effectively identical quantitative results in the statistics. Here, we compute instantaneous $c$-fields using the latter method. These fields are then used to estimate the values of $\mu$ and $\rho$ required for calculation of $\langle IV \rangle$ using a laminar flamelet assumption, i.e. that $\mu$ and $\rho$ are related to $c$ in the same manner as in a laminar flame.

\subsection{Uncertainty in reported quantities}
The main sources of uncertainty in mean quantities (i.e. terms in Eq.~\ref{e:RAETE}) stem from measurement noise, under-resolution in the velocity fields, and statistical convergence. In this study, uncertainty in $\left\langle L\right\rangle$, $\left\langle I\right\rangle$ and $\left\langle II\right\rangle$ were estimated by assuming a Gaussian process. The uncertainty in the baroclinic torque ($\left\langle III\right\rangle$) was determined by summing uncertainties in quadrature.

Uncertainty in $\left\langle IV\right\rangle$ is controlled by under-resolution of the velocity field, which results in a systematic bias towards smoother velocity gradients. The effects of under-resolution were evaluated in \citet{Kazbekov2019} by performing the complete analysis with different spatial resolutions (i.e. final interrogation box sizes). It was found that $\left\langle L\right\rangle$, $\left\langle I\right\rangle$, and $\left\langle II\right\rangle$ were unaffected by the interrogation box size, when comparing a larger interrogation box to that used in the reported data. However, this did not hold for $\left\langle IV\right\rangle$ across all the cases.

%The resolution of the velocity measurement has a direct effect on the magnitude of under-resolution due to the use of multi-point gradient estimators; however, it is limited by particle seeding density in the flow. The size of the interrogation window, used in TPIV cross-correlation, was varied to maximize the velocity resolution and minimize the reduction in measured gradients while maintaining low levels of measurement noise for the least turbulent case \cite{Kazbekov2019}. 

Calculation of $IV$ requires calculation of the third spatial derivative of velocity, which tends to amplify noise considerably. While $\left\langle IV \right\rangle$ was unaffected by the velocity resolution for Case~1, it was affected in the more turbulent Cases 2 and 3; there was a systematic increase in magnitude of the measured $\left\langle IV \right\rangle$ with decreasing interrogation box size, indicating that this term was under-estimated. Under-estimation of $|\langle IV \rangle|$ results in a corresponding under-estimation of $|\langle III \rangle|$. The bias uncertainty for Cases~2 and 3 was estimated based on comparison with (the fully converged) Case 1, as detailed in \citet{Kazbekov2019}. This is represented by asymmetric error bars in the corresponding plots.

\section{Results and Discussion}

\subsection{Conditioning on mean and instantaneous progress variable separately}
\begin{figure}[ht]
    \centering
    \subfigure[$\langle \mathrm{Terms} \, | \, c\rangle$ - Case 2]{\includegraphics[trim = {15 0 35 0},clip,width = 67 mm]{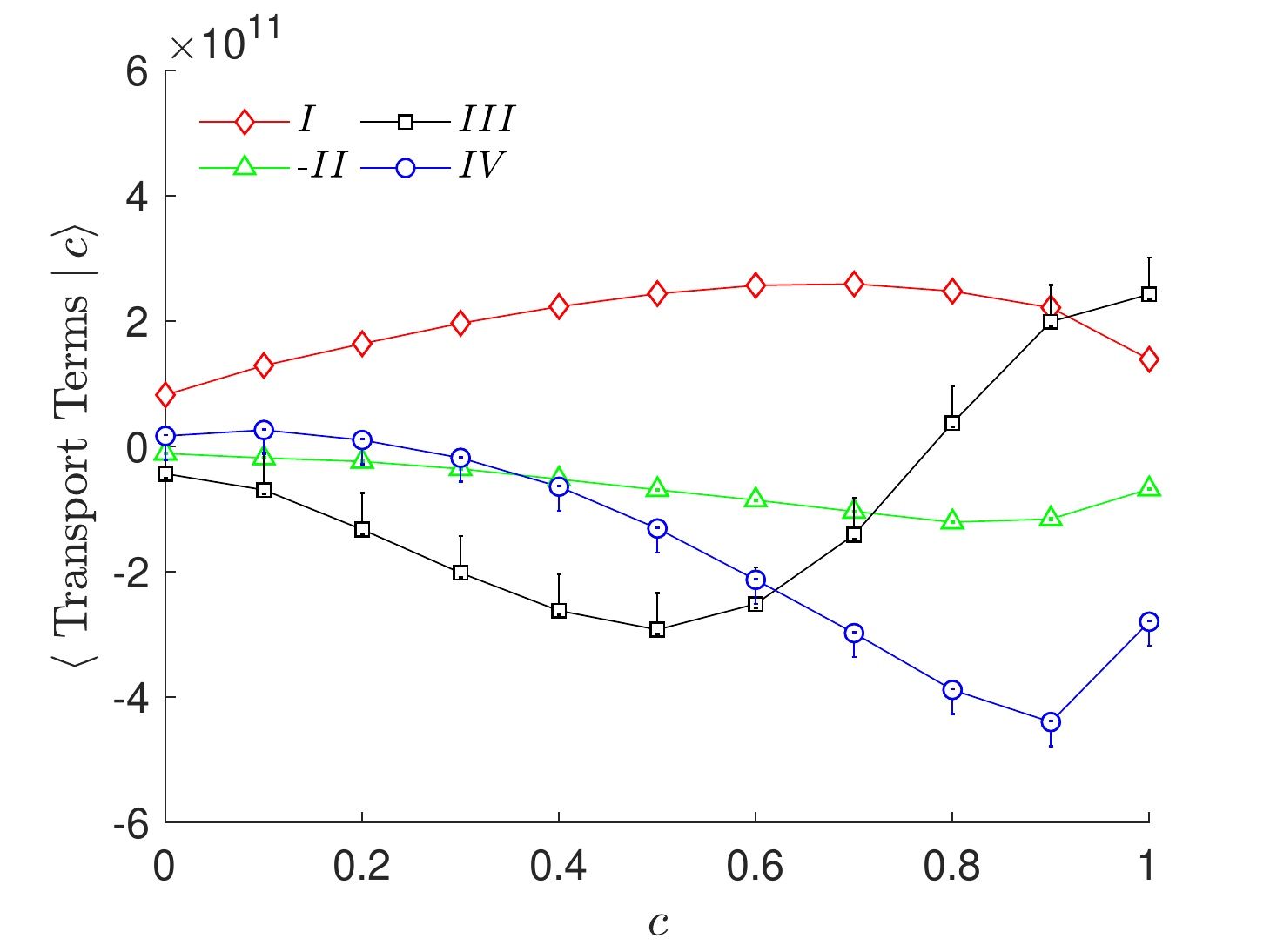}}
    \subfigure[$\langle \mathrm{Terms} \, | \, \cBar\rangle$ - Case 2]{\includegraphics[trim = {15 0 35 0},clip,width = 67 mm]{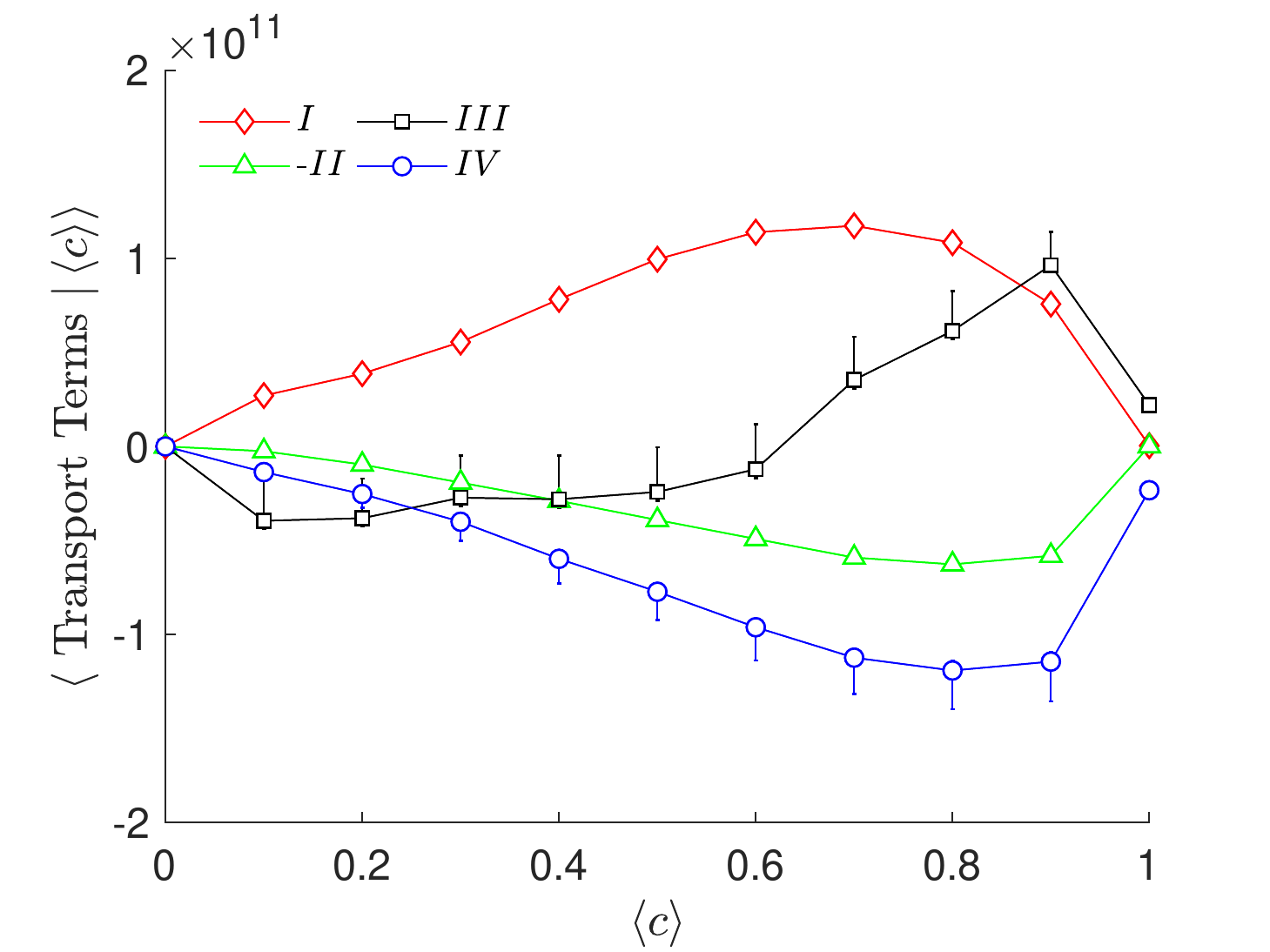}}
    \caption{Mean enstrophy transport terms conditioned on the instantaneous and mean progress variable for all cases.}
   \label{f:AllTermsCase2}
\end{figure}

We now compare the mean enstrophy budgets, as characterized by $\langle \mathrm{Terms} \, | \, \cBar \rangle$ and $\langle \mathrm{Terms}\, | \, c\rangle$; Fig.~\ref{f:AllTermsCase2} shows these terms for Case~2. Vortex-stretching is an important source of enstrophy, while dilatation and viscosity act to reduce enstrophy. Baroclinic torque, in contrast, can be either a sink or a source. It is noted that the skewness of the $\langle IV\rangle$ profiles towards higher $c$ and $\cBar$ is due to the increased temperature, and hence increased viscosity, in the products.

When conditioned on the instantaneous $c$, baroclinic torque acts as an enstrophy sink for $c\lesssim 0.7$ and a source for $c\gtrsim 0.7$. This is consistent with previous results that demonstrated attenuation of reactant-side vorticity and generation of counter-rotating product-side vorticity in flame/vortex interactions~\cite{Mueller1998, Louch1998}. The enstrophy attenuation by baroclinic torque is less pronounced when conditioned on the position within the mean flame-brush (\cBar), but there remains a region of strong enstrophy production due to baroclinic torque towards the product-side of the flame brush.

%Moreover, enstrophy attenuation is more pronounced when conditioned on $c$. With the exception of baroclinic torque, the sign of each term is maintained in both plots, but the magnitude of the mean terms is greater when conditioned on instantaneous $c$. 

%Dilatation is strictly negative and similar when conditioned on $c$ and $\cBar$ -- in both cases, it is skewed towards higher values of $c$ and $\cBar$. The profiles of $\langle I \rangle$, and $\langle IV \rangle$ differ when conditioned on $\cBar$ and $c$. Vortex-stretching across the instantaneous flame is more uniform than across the flame brush; however, both peak at approximately the same value of the conditioning variable (i.e. $c\approx0.7$). The difference between these profiles arises due to the movement of the flame in physical space. The flame brush physically overlaps with the inner shear layer, thus, $\langle I \, | \, \cBar \rangle$ is expected to be similar to that in the shear layer. Conversely, the instantaneous premixed flame is thinner than the flame brush and resides predominantly within the shear layer. This results in a more uniform $\langle I \, | \, c\rangle$ profile. 

\begin{figure}[ht]
    \centering
     \subfigure[$\langle III \, | \, c\rangle$ and $\langle IV \, | \, c\rangle$]{\includegraphics[trim = {15 0 35 0},clip,width = 67 mm]{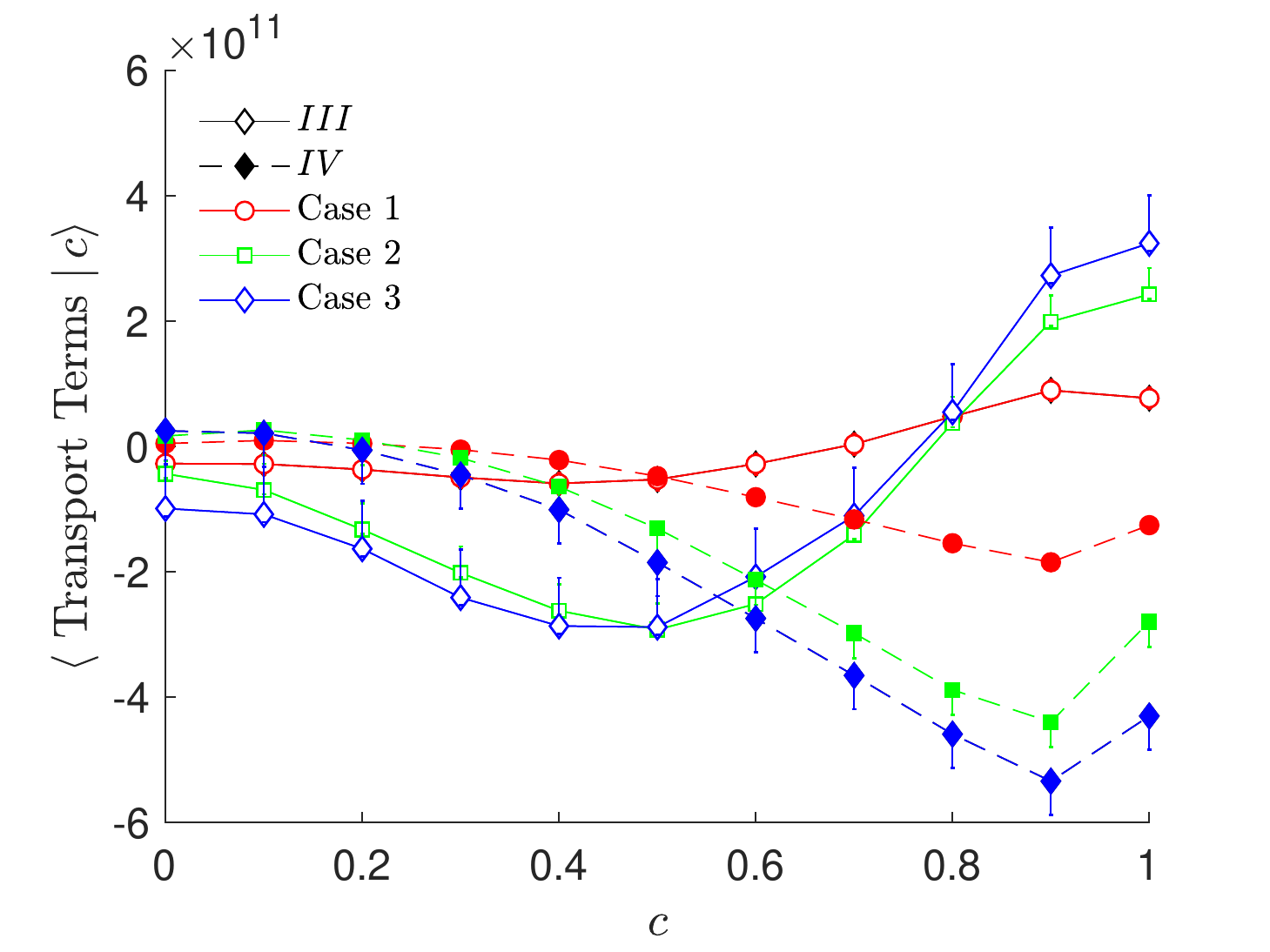}}
    \subfigure[$\langle III \, | \, \cBar\rangle$ and $\langle IV \, | \, \cBar \rangle$]{\includegraphics[trim = {15 0 35 0},clip,width = 67 mm]{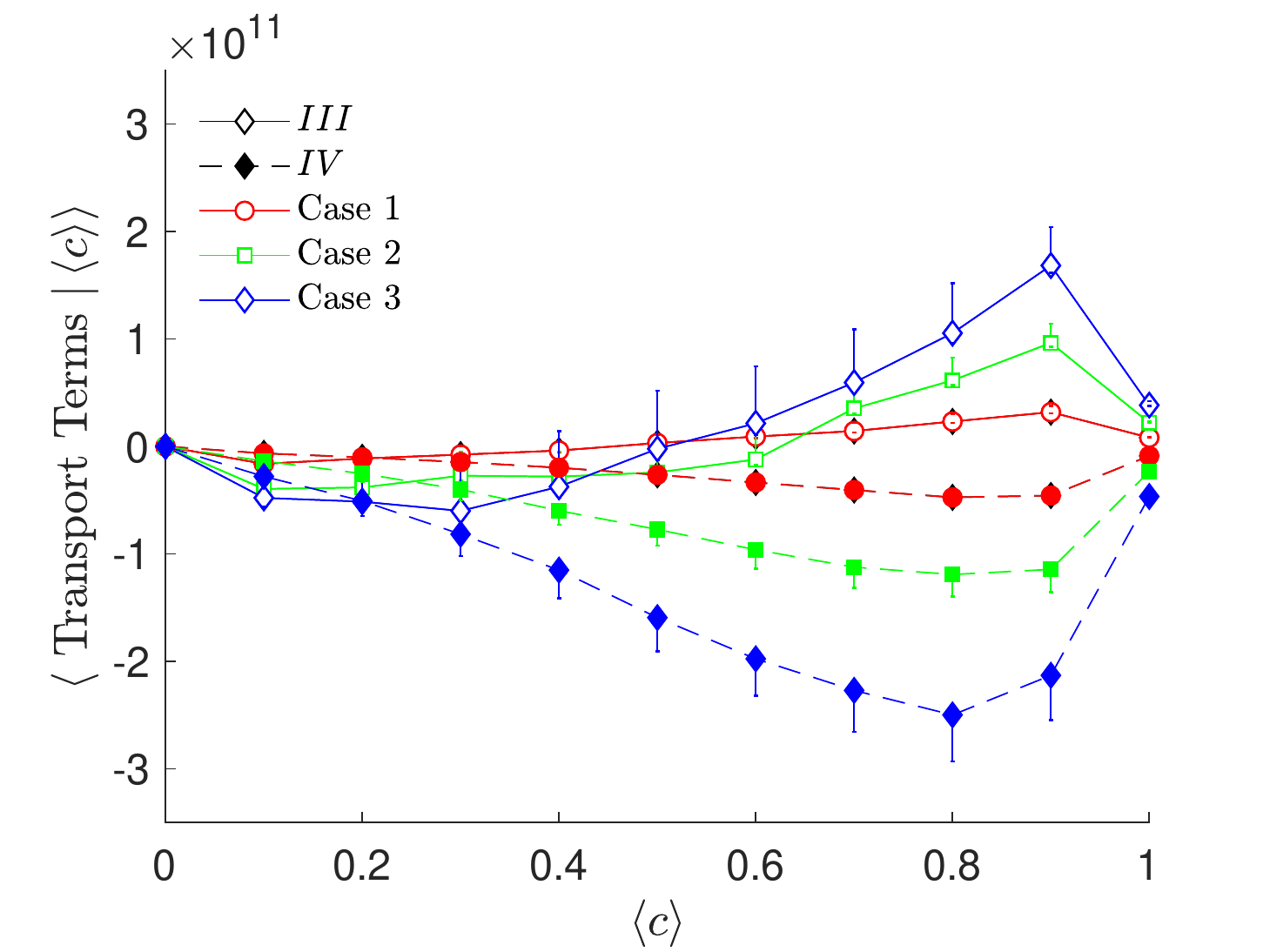}}
    \caption{Mean baroclinic torque and viscous diffusion conditioned on the instantaneous and mean progress variable across all cases.}
    \label{f:IIIandIV}
\end{figure}

%In Fig.~\ref{f:inst_c}, dissipation of enstrophy by viscous action and diffusion is significantly skewed towards the products across the flame front, whereas the skewness is not as pronounced across the flame brush. The increase in temperature and temperature-dependent viscosity across the flame results in enhanced dissipation, especially within the heat-release zone of a premixed flame. The $\langle IV \, | \, \cBar\rangle$ is similarly skewed towards the hot products, but the valley is wider due to correspondingly broadened flame brush, caused by physical movement of the flame within the shear layer. 

%Vorticity attenuation by baroclinic torque at the early stages, followed by strong vorticity production near the reaction zone was reported by \citet{Louch1998} and \citet{Mueller1998}. Moreover, they also found that the vorticity attenuation through baroclinicity was generally same or greater than vorticity production. A similar trend is seen in Fig.~\ref{f:inst_c}~(b) for the instantaneously conditioned baroclinic torque. However, little to no baroclinic enstrophy attenuation is observed in the flame brush, albeit the vorticity production is similar to $\langle III \, | \, c\rangle$. 

Figure~\ref{f:IIIandIV} shows the variation of $\langle III \, | \, \cBar \rangle$ and $\langle III \, | \, c\rangle$ for all three cases; the viscous term also is included for reference. Increasing the flow rate, and hence increasing both the turbulence intensity and mean pressure gradient, simultaneously increases the magnitude of both the baroclinic torque and viscous terms in the enstrophy transport equation. Hence, flame-induced enstrophy transport is significant in all flames studied here. 

%The influence of increasing the mass flow rate, and correspondingly the turbulence intensity, results in stronger enstrophy production by baroclinic torque within the flame brush, but attenuation remains small. On the other hand, both production and atttenuation by baroclinic torque increase in magnitude across the instantaneous flame, represented by $\langle III \, | \, c\rangle$. 

%The difference in enstrophy attenuation is related to the source of pressure gradient in the flow. Mean swirling flow induces radial and axial pressure gradients that remain constant in statistically stationary turbulent flows. Their effect is observed in the $\cBar$-conditioned profiles. Alternatively, pressure fields induced by turbulent vortices and the flame have a direct influence on the immediate baroclinic behaviour which may change as the turbulent eddies transit the flame front. These effects manifest themselves in the $c$-conditioned profiles. The data thus demonstrates that the local turbulence conditions and thermodynamic properties of the fluid have a strong influence on attenuation of enstrophy, but enstrophy production is more affected by the pressure gradient imposed by the mean swirling flow. 

\subsection{Multiply-conditioned analysis}
While the above results demonstrate the importance of baroclinic torque on enstrophy transport, it is difficult to deduce the relative impacts of local turbulence and the large-scale pressure field. Here, we attempt to observe these effects by simultaneously conditioning the data on $c$ and \cBar. If only the local turbulence-flame interactions were significant, the baroclinic torque should depend only on $c$ and not on $\cBar$; profiles conditioned on different $\cBar$ should collapse. If the large-scale pressure field was dominant, the variations with $c$ should be small compared to the variations with $\cBar$.

%\begin{figure}[ht]
%    \centering
%    \includegraphics[width=67mm]{III_cbar_map}
%%    \caption{A map of $\langle III \, | \, c, \cBar \rangle$ for Case~2.}
 %   \label{f:c_map}
%\end{figure}

\begin{figure}[!ht]
    \centering
    \subfigure[Case 1]{\includegraphics[trim = {5 0 35 0},clip,width = 67 mm]{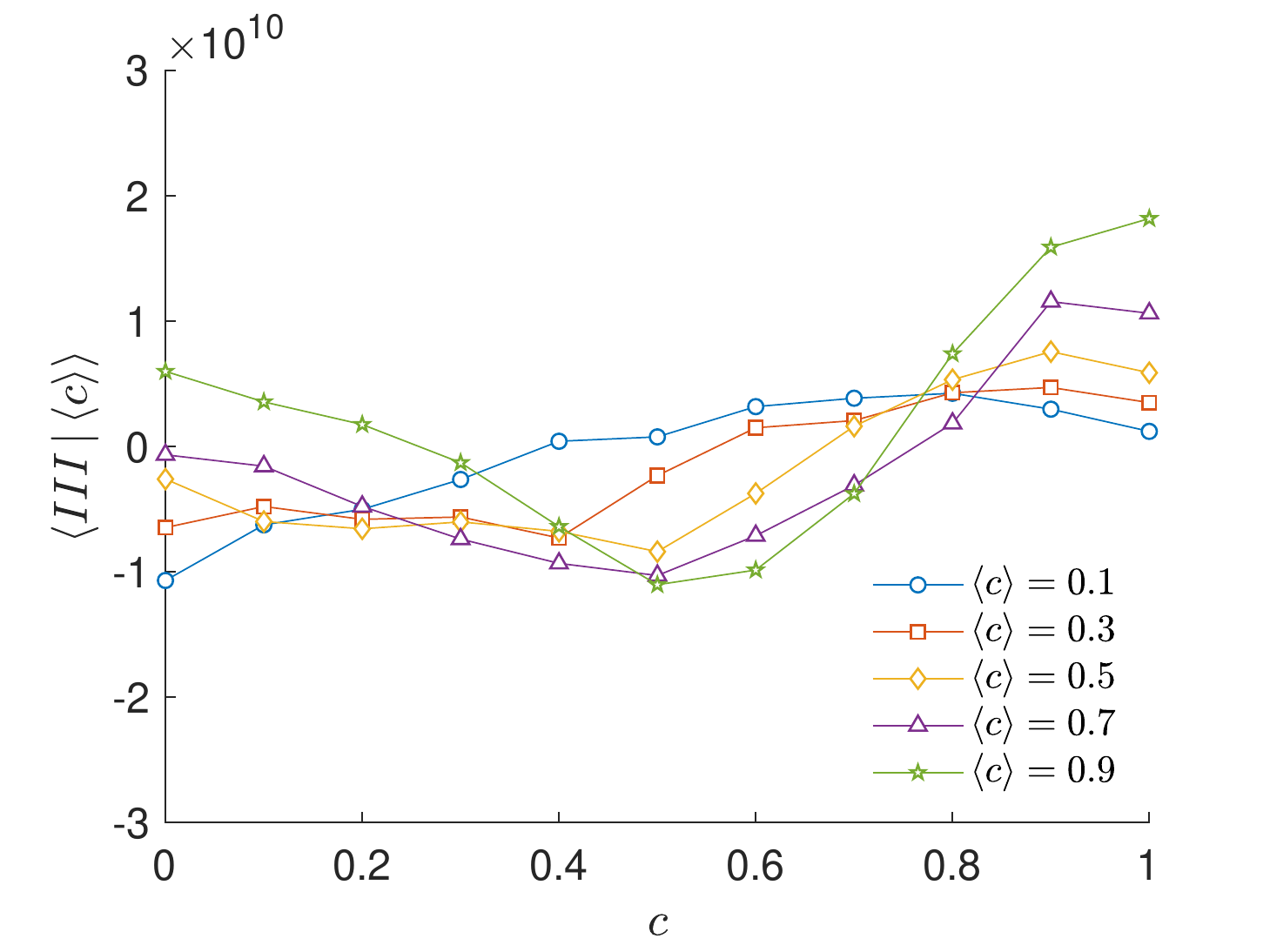}}
    \subfigure[Case 2]{\includegraphics[trim = {5 0 35 0},clip,width = 67 mm]{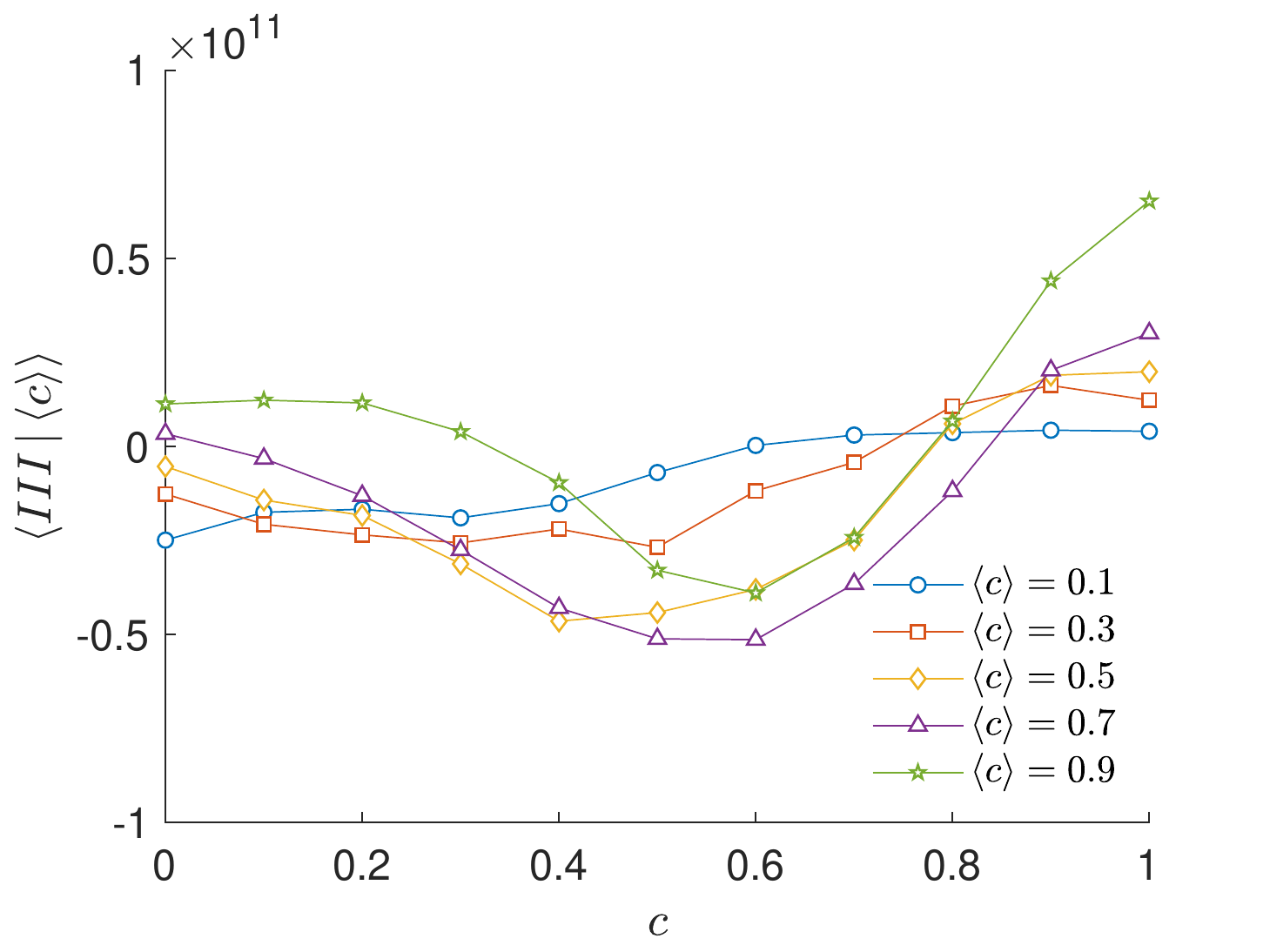}}
    \subfigure[Case 3]{\includegraphics[trim = {5 0 35 0},clip,width = 67 mm]{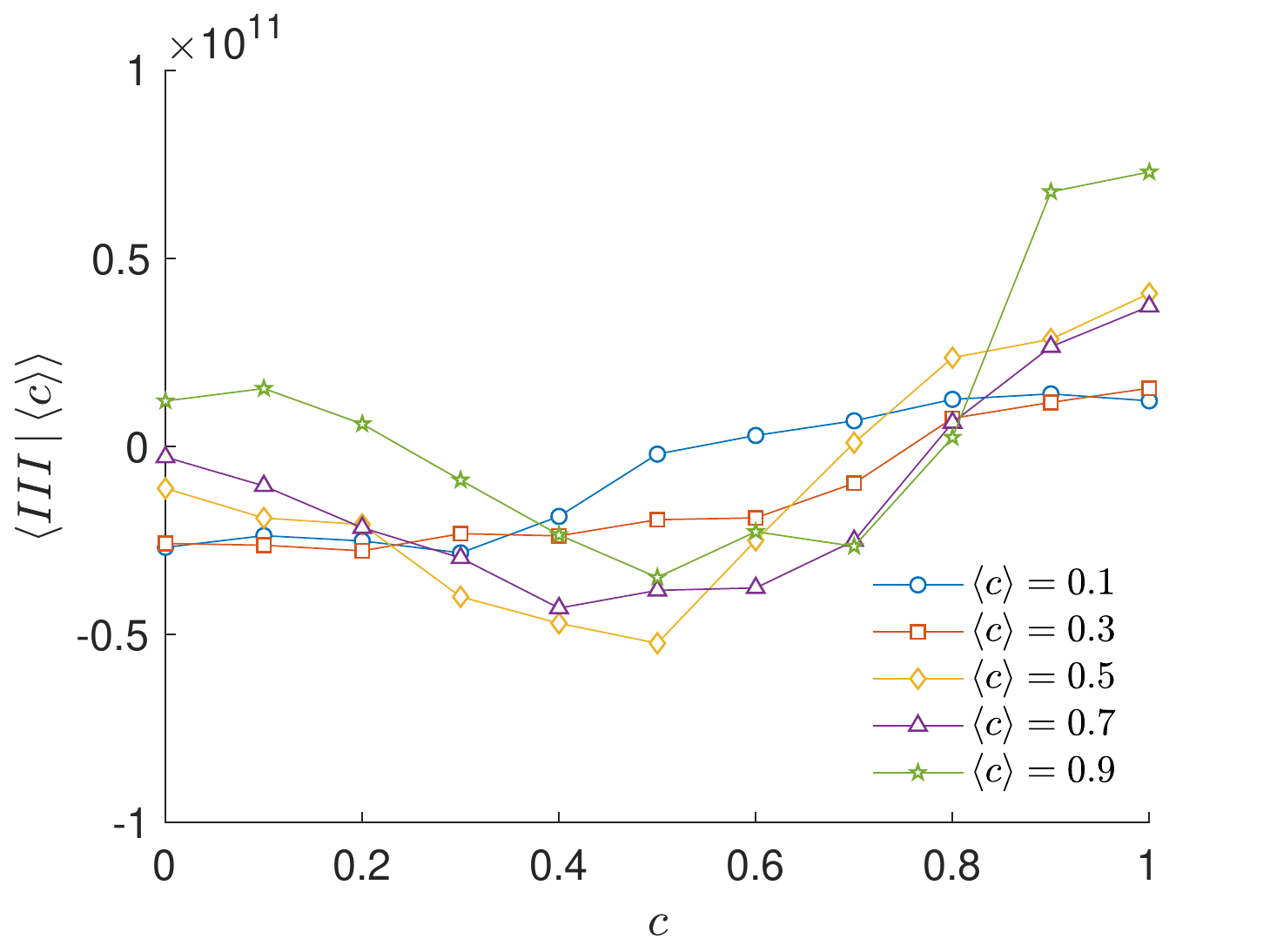}}
    \caption{Mean baroclinic torque versus instantaneous progress variable, conditioned on the mean progress variable.}
    \label{f:double_cond}
\end{figure} 

Figure~\ref{f:double_cond} shows $\langle \mathrm{III} \, | \, c, \cBar\rangle$ for all three cases. Profiles are provided as functions of $c$ at different values of $\cBar$. The uncertainty bars have been omitted for clarity, but are similar to those in Figs.~\ref{f:AllTermsCase2} and \ref{f:IIIandIV}. %Figure~\ref{f:c_map} shows the 2D profile of $\left\langle III|c,\cBar\right\rangle$ versus $c$ and $\cBar$ for Case 2.

The main takeaway from Fig.~\ref{f:double_cond} is that baroclinic torque is a function of both position within the flame ($c$) and position within the combustor (as characterized by \cBar). For example, enstrophy production by baroclinic torque at high values of $c$ increases at high values of $\cBar$. Similarly, enstrophy attenuation at low-and moderate values of $c$ generally increases with $\cBar$, but seems to peak at $\cBar\approx 0.7$. This may be due to the low occurrence of low $c$ at high $\cBar$. At a given value of $\cBar$, baroclinic torque is a function of $c$ across all cases. Hence, both local turbulence/flame interactions and the global flow structure/pressure are significant for determining the baroclinic torque -- and thus the enstrophy transport -- in the swirl flames studied here.

\subsection{Effect of axial location}
In the above analysis, we have equated the position of the fluid within the global pressure field with its position within the \cBar-field. This is partially justified by the overlap between the flame brush and shear layer separating the inflowing swirling reactants from the CRZ, as shown in Fig.~\ref{f:setup}; the \cBar-field is at a relatively fixed radial position relative to the global flow structure. However, the flow also is inhomogeneous in the axial direction, which can have an effect on the enstrophy transport.

\begin{figure}[!t]
    \centering
    %\subfigure[$\cBar = 0.1$]{\includegraphics[width = 67 mm]{III_triple_cond_C2_01}}
    \subfigure[$\cBar = 0.5$]{\includegraphics[trim = {5 0 35 0},clip,width = 67 mm]{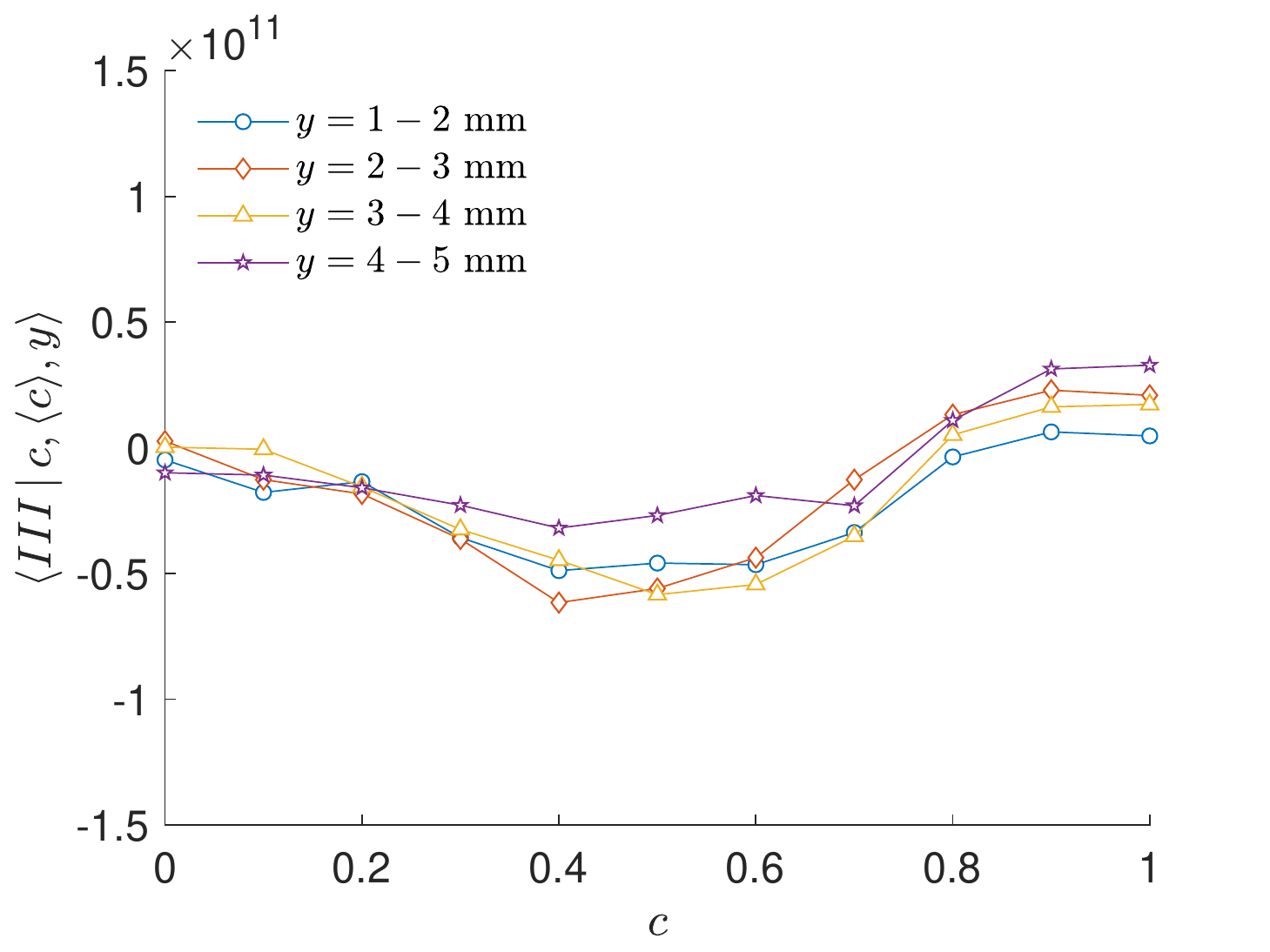}}
    \subfigure[$\cBar = 0.9$]{\includegraphics[trim = {5 0 35 0},clip,width = 67 mm]{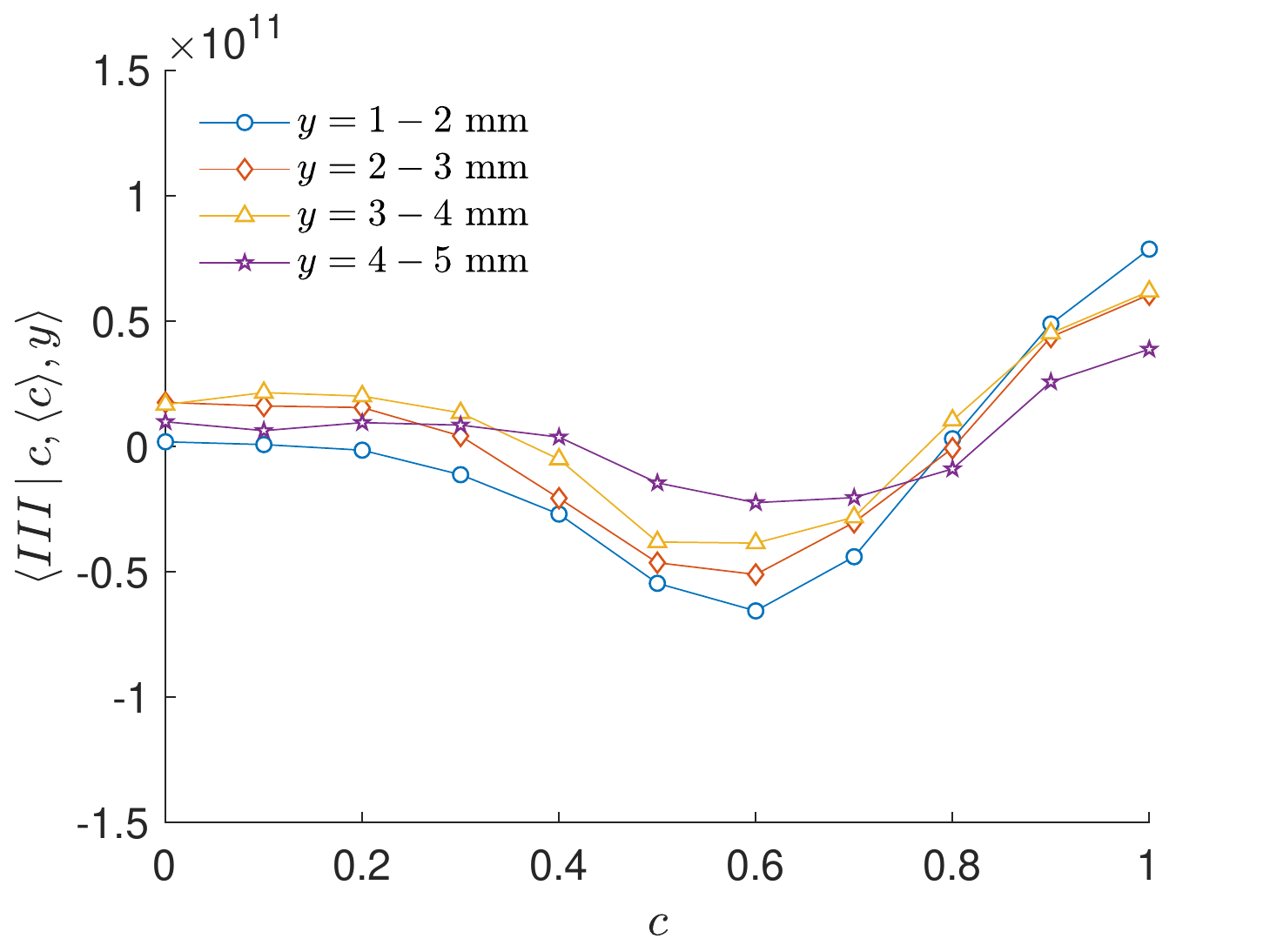}}
    \caption{Mean baroclinic torque conditioned on position within the instantaneous flame, position within the flame brush, and axial location within the combustor for Case 2.}
\label{f:triple_cond}
\end{figure}

This is demonstrated in Fig.~\ref{f:triple_cond}, which shows $\langle III \, | \, c, \cBar, y \rangle$ for Case 2 at $\cBar = 0.5$, and $0.9$. Axial bands of 1~mm height are used to condition the data on the height above the burner. 

The baroclinic torque changes with axial position in a manner that depends on the position through the flame brush. For example, at $\cBar=0.9$, the region near the combustor exit has the highest magnitude attenuation and generation of enstrophy by baroclinic torque. The magnitudes steadily decrease with downstream distance. This may be due to the strong pressure gradients near the inner shear layer between the inflowing reactants and CRZ (i.e. at high $\cBar$) near the nozzle exit.

In contrast, at $\cBar = 0.5$, the most upstream location has the weakest positive baroclinic torque. The most downstream location has the least attenuation and the strongest enstrophy production. These results further demonstrate the complicated interactions between large-scale flow features and the flame in the context of flame-induced vorticity transport.

%The variation in the axial location is observed to have minimal effect on the baroclinic torque. The largest variation is observed at $\cBar=0.9$, where the term $III$ is strongest 0-2~mm above the nozzle, followed by drop in intensity at higher $y$. This suggests that the mean axial pressure gradient has a small effect on baroclinic torque activity.

\section{Conclusion}

This study has experimentally examined the enstrophy production and attenuation due to baroclinic torque in turbulent premixed swirl flames using high-resolution TPIV and \chem{CH_2O} PLIF. The data allowed calculation of the ensemble mean baroclinic torque, conditioned on the instantaneous progress variable, mean progress variable, and axial position in the combustor. The data demonstrate that (i) significant vorticity attenuation and generation occurs in swirl flames due to baroclinic torque across the range of conditions studied; (ii) both large-scale (associated with the global flow field) and small-scale (associated with the local turbulence) pressure gradients affect the baroclinic torque, and (iii) the strongest flame-induced vorticity production occurs at high values of $c$, high values of \cBar, and near the nozzle of the burner. Hence, the combined actions of the flame, turbulence, and swirling flow have a large impact on the turbulence dynamics -- particularly on flame-scale turbulence production -- in the swirl combustor studied here. Future work should examine the impact of this flame-scale turbulence production on the turbulent burning rate and inter-scale interactions. Furthermore, efforts should be made to extend the analysis to other conditions involving strong large-scale pressure fields, such as high-pressure swirl flames and high-speed combustion.

\section*{Acknowledgments}
\label{Acknowledgments}
This work was supported by the US Air Force Office of Scientific Research under grant FA9550-17-0011, Project Monitor Dr. Chiping Li. A. Kazbekov acknowledges the support of the NSERC-PGS D fellowship.

%% References can be added with or without bibTeX database
%% Note that the PROCI references style is considered Elsevier non-standard.
%% The original Elsevier bibliography style, elsarticle-num.bst prints paper titles as part of the references, which is different from 

\bibliography{symp2020ref} %%User-specified
\bibliographystyle{elsarticle-num-PROCI}

% \begin{thebibliography}{99}
% \bibitem{Westbrook_1984} C. Westbrook, F. Dryer, Progress in Energy and Combustion Science 10 (1984) 1--57.
% \bibitem{Peters_2002} N. Peters, G. Paczko, R. Seiser, K. Seshadri, Combustion and Flame 128 (2002) 38--59.
% \end{thebibliography}

\end{document}